\documentclass[a4paper]{article}
\usepackage{amssymb,amsmath,cite,citesort,graphicx}
\newcommand{\gref}[1]{(\ref{#1})}
\newcommand{\td}{\text{d}}

\newcommand{\de}[2]{\frac{\td #1}{\td #2}}
\newcommand{\dde}[2]{\frac{\td^2 #1}{\td #2 ^2}}
\newcommand{\be}{\begin{equation}}
\newcommand{\ee}{\end{equation}}

\title{Diffusion of intrinsic localised modes by attractor hopping}
\author{Matthias Meister\footnote{email: matthias@unizar.es} \footnote{also at: Instituto de
Biocomputaci\'on y F\'\i sica de Sistemas Complejos, Universidad de Zaragoza, 50009 Zaragoza, Spain}
\\ Dpto. F\'\i sica de la Materia
Condensada \\ Facultad de Ciencias, Universidad de Zaragoza \\  50009 Zaragoza, Spain
\\[3mm] and \\[3mm]
Luis V\'azquez\footnote{also at: Centro de Astrobiolog\'\i a (CSIC-INTA),
28850 Torrej\'on de Ardoz, Spain}
\\
Dpto. Matem\'atica Aplicada \\ Facultad de Inform\'atica, Universidad Complutense de
Madrid \\ 28040 Madrid, Spain
}
\begin{document}
\maketitle
\abstract{
Propagating intrinsic localised modes exist in the damped-driven discrete sine-Gordon chain
as attractors of the dynamics. The equations of motion of the system are augmented with Gaussian
white noise in order to model the effects of temperature on the system. The noise induces random
transitions between attracting configurations corresponding to opposite signs of the propagation
velocity of the mode, which leads to a diffusive motion of the excitation. The Heun method is used
to numerically generate the stochastic time-evolution of the configuration. We also present a
theoretical model for the diffusion which contains two parameters, a transition probability $\Theta$
and a delay time $\tau_{A}$. The mean value and the variance of the position of the intrinsic
localised mode, obtained from simulations, can be fitted well with the predictions of our model,
$\Theta$ and $\tau_{A}$ being used as parameters in the fit. After a transition period following the
switching on of the noise, the variance shows a linear behaviour as a function of time and the mean
value remains constant. An increase in the strength
of the noise lowers the variance, leads to an increase in $\Theta$, a decrease in $\tau_{A}$ and
reduces the finite average distance a mode travels during the transition period.
}
%
%
\fontsize{12}{16}\selectfont
\section{Introduction}
Intrinsic localised modes (ILMs) are localised excitations that can occur in a system due to the
interplay of discreteness and nonlinearity, for instance see \cite{DisNon}.
These modes exhibit internal dynamics, like oscillations
or rotations of constituents of the system in the localisation region.
Therefore, these modes also are known as discrete breathers, alluding to the superficial similarity
between them and the breather solutions of the continuum sine-Gordon equation. The majority of the
literature deals with strictly time-periodic, non-propagating ILMs.
 For this specific type of discrete breather there are rigorous proofs of
existence for the conservative \cite{MaAu94,Au97} and dissipative \cite{Sep97} cases. From the
proofs it is clear that the requirements for the existence of these modes are rather weak, which
makes ILMs quite generic, in contrast to breather solutions in the continuum. An overview of some
recent results can be found in \cite{Mart03}. \newline
First predictions
\cite{Siev88} of the existence of such modes, however, did not exclude the case of propagating
discrete breathers and numerical results \cite{Tak90} soon showed the existence of mobile ILMs.
Several aspects of such travelling modes have been investigated so far, for example the interaction
with an impurity \cite{Cuev02-1} or the effects of a bending of the chain along which the excitation
is travelling in connection with models of DNA \cite{Cuev02-2,Cuev02-3}.
To our knowledge, the diffusion of intrinsic localised modes has not been looked at yet;
this is in stark contrast to the thoroughly investigated topic of soliton / solitary wave diffusion.
For the latter, see for instance \cite{Pas85,Rod90,Phi4,Qui99,Qui00,Mei01} and references therein.
\newline
We study this problem in the discrete sine-Gordon chain (also known as the standard
Frenkel-Kontorova model) under damping and harmonic driving; the equations of motion, in
dimensionless units, are: \be
\label{Eqmotion}
\dde{u_{n}}{t}=-\frac{1}{2\pi}\sin(2\pi u_{n}) + C(u_{n+1}+u_{n-1}-2u_{n})-\alpha \de{u_{n}}{t}
+F\sin(\omega_{0}t).
\ee
This equation describes the dynamics of a chain of identical damped and driven pendula
in a homogeneous gravitational field, with harmonic nearest neighbour coupling. The quantity $u_{n}$
corresponds to the angle of deviation (in units of $2\pi$) of the $n$-th pendulum from the position
of minimum energy. Certain systems of Josephson junctions can also be described by equation
\gref{Eqmotion}, see \cite{JJA} for example.
\newline
Numerical
results \cite{Mar01} show that intrinsically localised modes exist as attractors of the dynamics
governed by this equation. In particular, within certain ranges of the parameters $C,\alpha,
\omega_{0}$ and $F$ the attracting configurations are propagating ILMs with well defined propagation
velocities, dependent on the system parameters; apart from the fact that the discrete sine-Gordon
equation is well-known, this has been the main motivation for our choice of equation
\gref{Eqmotion}.
Due to the symmetry in the equations of motion, for
each absolute value of the velocity we can have positive and negative sign, i.e. propagation to the
right or left, respectively. There exist regions in parameter space where more than one absolute
value of the propagation velocity $v$ is possible at fixed parameters. \newline
In order to model temperature we add Gaussian white noise
terms $\xi_{n}(t)$ of strength $\sigma$, i.e.
\begin{eqnarray}
\langle \xi_{n}(t) \rangle &= 0 \nonumber \\
\langle \xi_{m}(t)\xi_{n}(t^{\prime}) \rangle &= \sigma^{2}\delta_{mn}\delta(t-t^{\prime})
\end{eqnarray}
to the equations of motion \gref{Eqmotion} for the individual pendula. Such noise terms can be
described in a clearer way by Wiener processes \cite{Gard}, but as the theoretical treatment of the
diffusive motion we present in sections \ref{secFirst} and \ref{secImpr} does not make use of the
formalism of stochastic differential equations, we do not tarry here. The relation between $\sigma$
and the temperature $T$ is $\sigma^{2}=2\alpha k_{B} T$, where $k_{B}$ is Boltzmann's
constant.
It turns out that the noise causes random transitions between basins of attraction corresponding to
opposite signs of the propagation velocity of the discrete breather, which brings about the
diffusive motion of the excitation. \newline
In section \ref{secNum} we present all our
numerical results, including the comparison with predictions to be derived in the next two sections.
Section \ref{secFirst} discusses a theoretical model for the diffusive behaviour of the modes which
uses only the transition probability $\Theta$ between basins of attraction as a parameter; it is
then improved in section \ref{secImpr} by the introduction of the delay time $\tau_{A}$ as a second
parameter. We discuss the results and draw some conclusions in section \ref{secDiss}.
An appendix lists some intermediate results related to section \ref{secImpr}.
\section{Numerical Results}
\label{secNum}
\begin{figure}
\begin{center}
\includegraphics{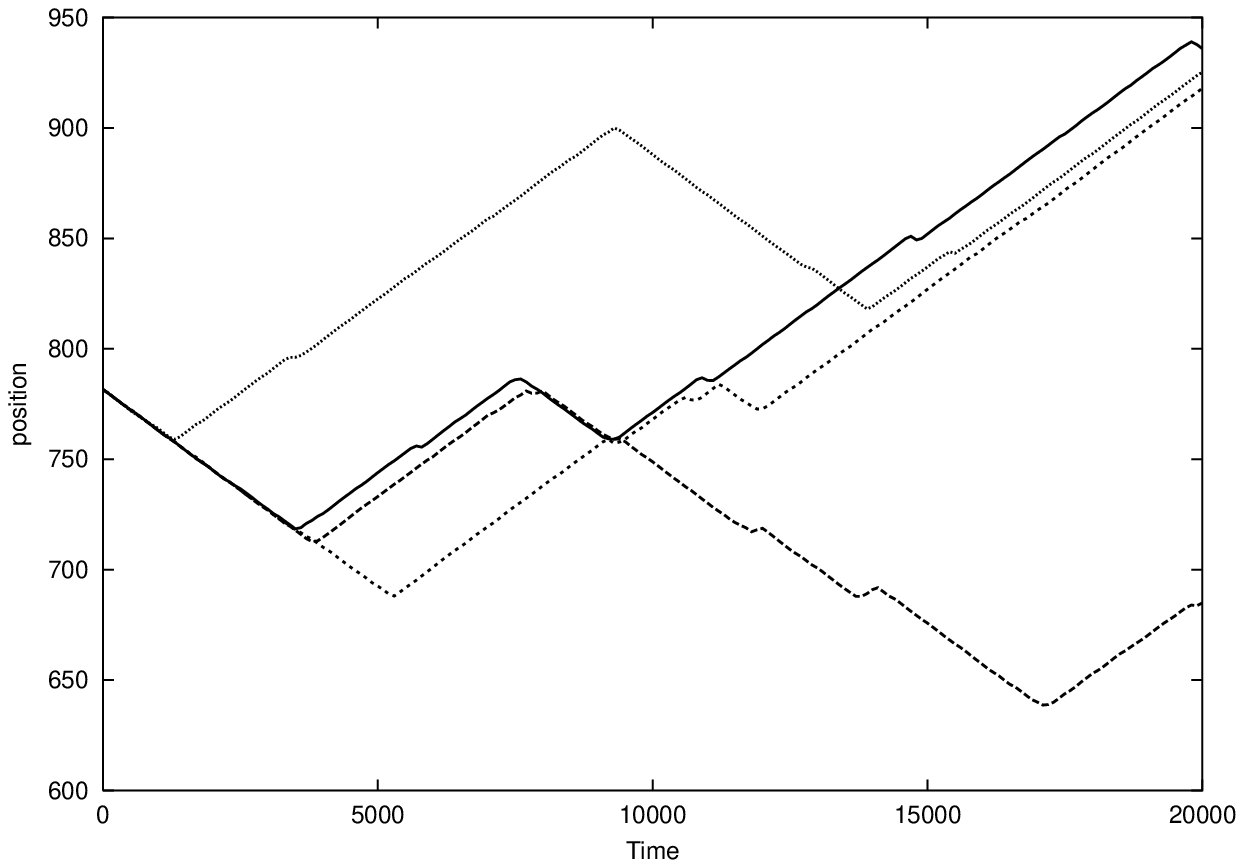} \includegraphics{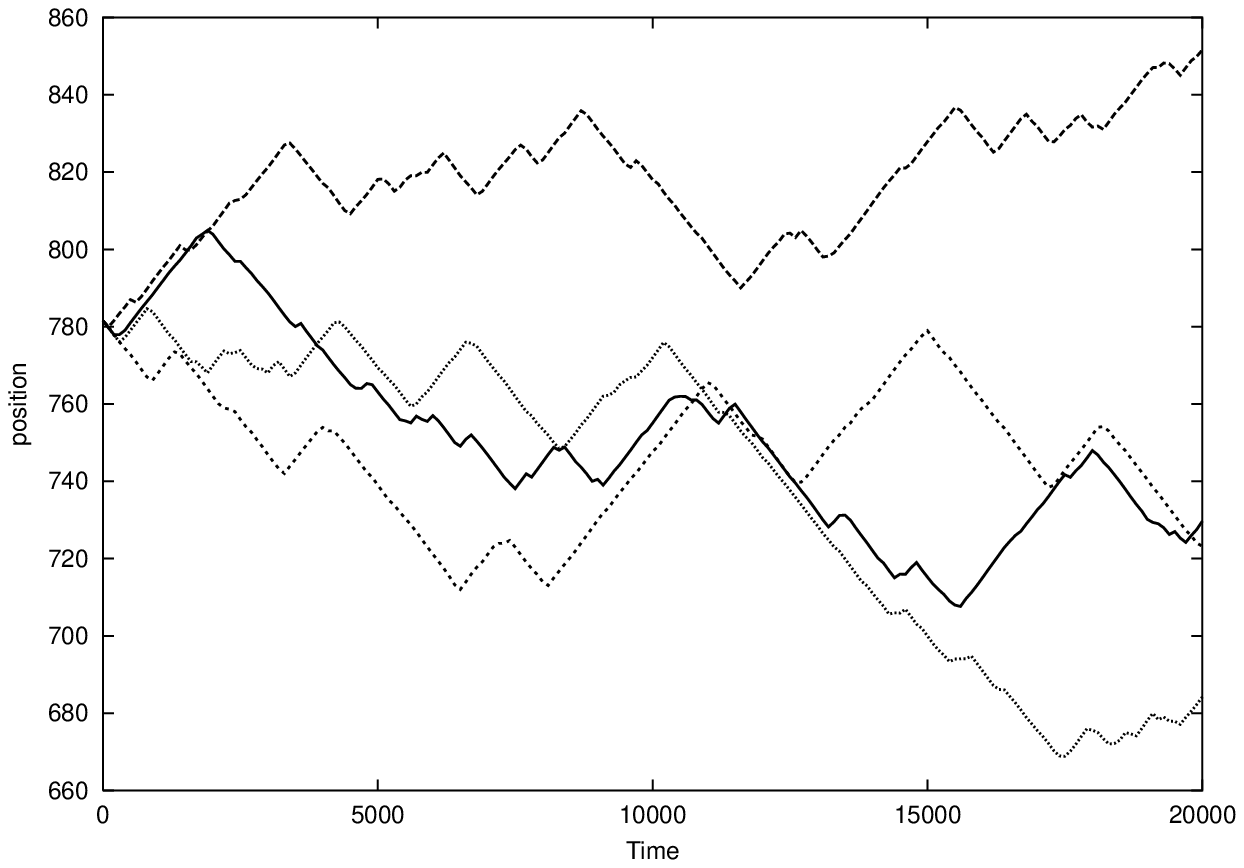}
\end{center}
\caption{Some sample trajectories for two different strenghts of the noise. Top: $\sigma=7 \cdot
10^{-5}$, Bottom: $\sigma=12 \cdot 10^{-5}$. The position of the localised mode has been calculated
every 100 time units.} \label{figTraj}
\end{figure}

The simulations are done in a chain of 1000 particles with periodic boundary conditions (i.e. a
ring). The stochastic time evolution according to \gref{Eqmotion} with the noise terms added is
generated by the Heun algorithm (time step $\Delta t=0.01$). We use parameter
values $\alpha=0.02, F=0.02, \omega_{0}=0.2\pi$ and $C=0.890$ along with several values of $\sigma$.
The local energies \be E_{n}=\frac{1}{2}p_{n}^{2}+\frac{1}{(2\pi)^{2}}\left[1-\cos(2\pi
u_{n})\right] \ee
are used to define the position $X$ of the localised mode as
\be
\label{XDef}
X=\frac{\sum\limits_{n=-2}^{n=+2} (n_{0}+n)E_{n_{0}+n}}{\sum\limits_{n=-2}^{n=+2}E_{n_{0}+n}}
\ee
where the site $n_{0}$ is chosen in the following way: The distribution of the $E_{n}$
sharply peaks around the location of the ILM. Usually, this peak has only one maximum. In this case,
$n_{0}$ is the site where this maximum occurs. Sometimes, the peak displays a structure
maximum-minimum-maximum on three subsequent sites. In this case $n_{0}$ is the site of the peak's
minimum. Using \gref{XDef} we calculate the position of the excitation out of the numerically
generated configurations. Averaging over 500 realisations, we obtain the time evolution of the mean
value of $X$ as well as the variance of the position.
In figure \ref{figTraj} we show some sample trajectories.
We can see that the noise causes the system to jump between attracting states corresponding to
opposite signs of the propagation velocity $v$. At the parameter values chosen, the system is known
\cite{Mar01} to allow for at least two values of $|v|$, which are approximately $0.0186$ and roughly
twice that. As the trajectories shown already hint and a closer quantitative look confirms, only the
former of the two values is involved. We cannot {\it strictly} exclude the appearance of more than
just this one value of $|v|$, but our numerical results show that if this happens, it is a rare
event. Transitions $v\approx 0.0186 \leftrightarrow v\approx -0.0186$ are clearly dominant.
\begin{figure}
\begin{center}
\includegraphics[scale=0.5]{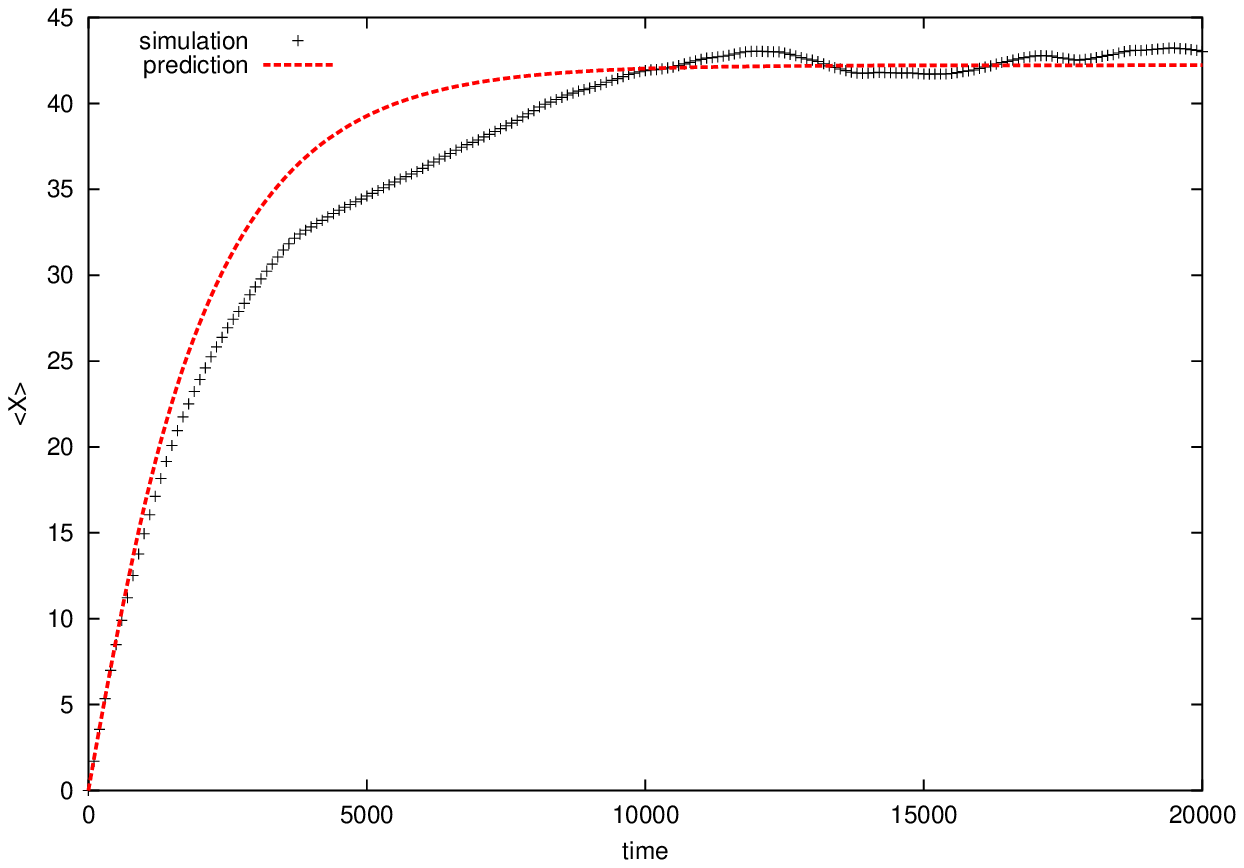} \includegraphics[scale=0.5]{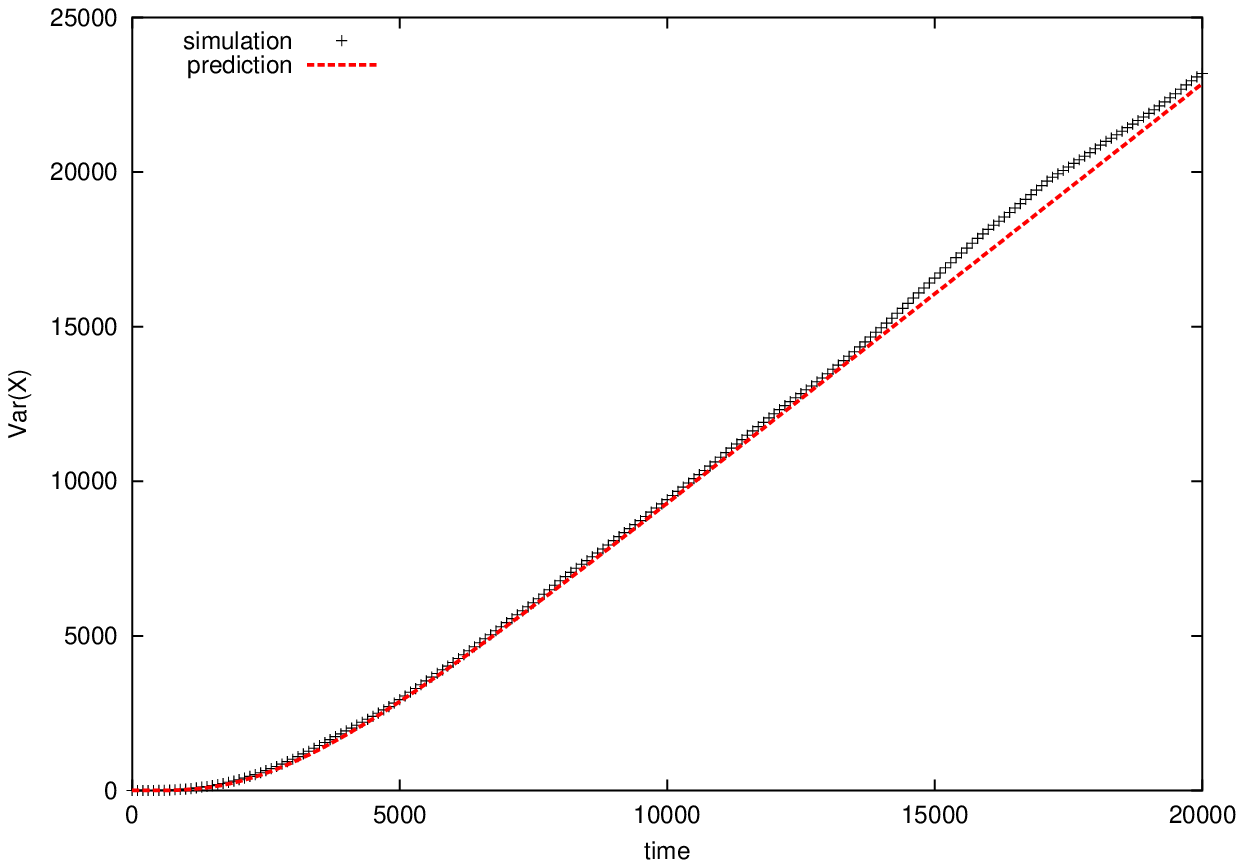}
\newline \includegraphics[scale=0.5]{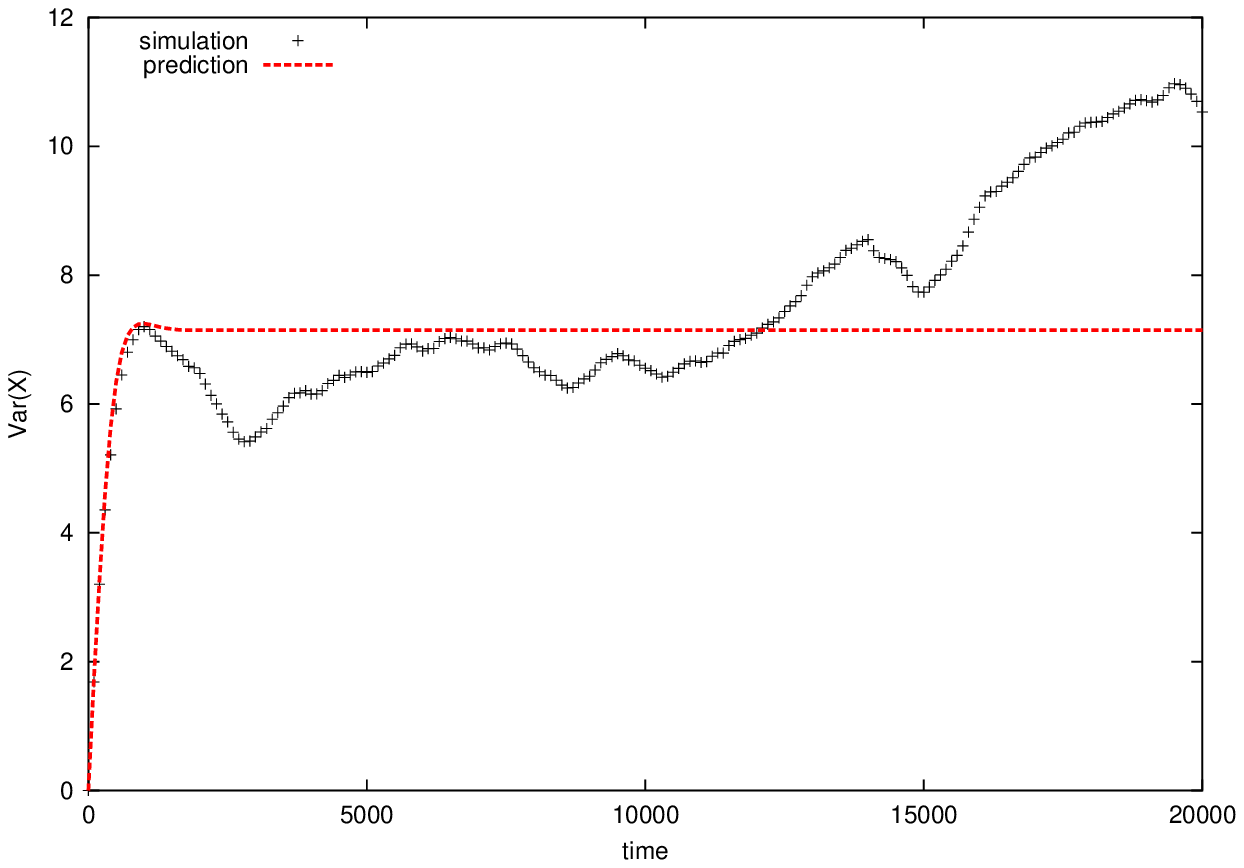}
\includegraphics[scale=0.5]{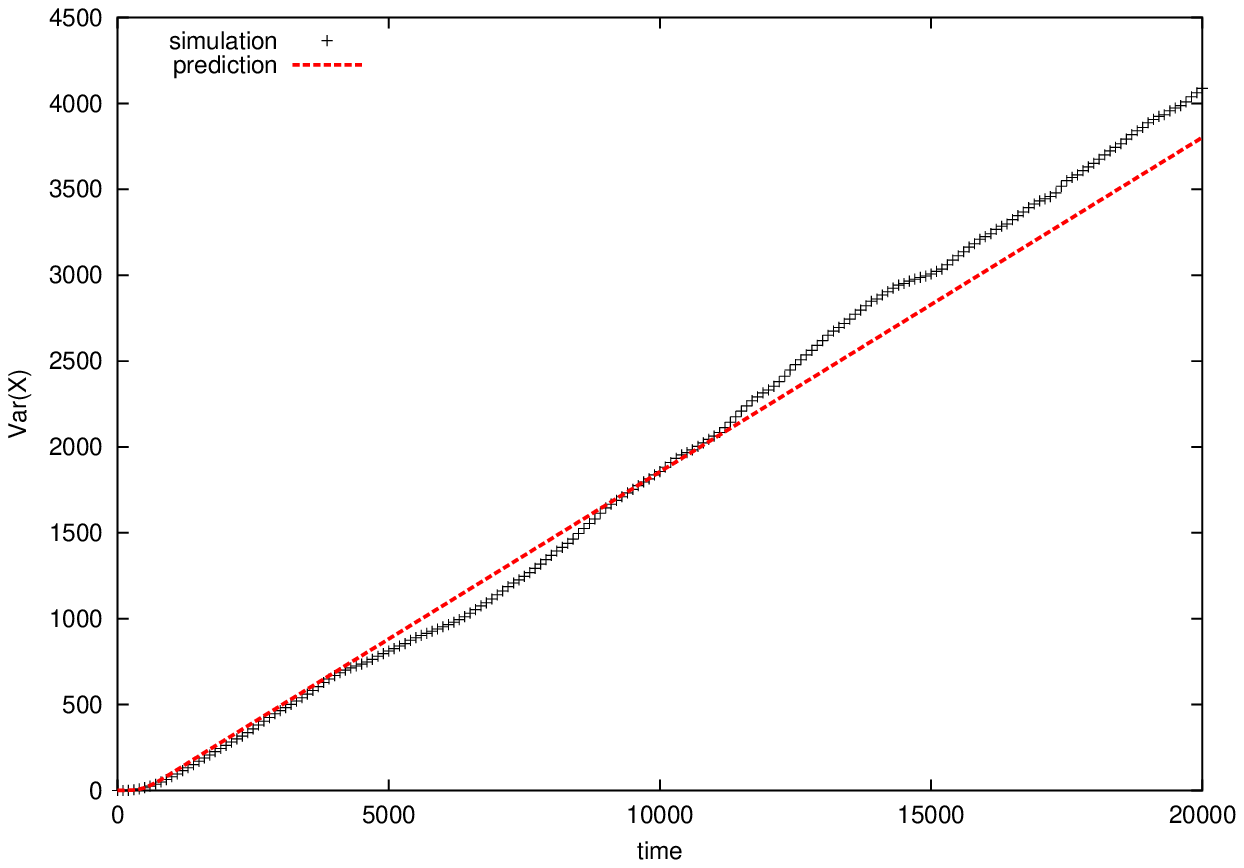} \newline \end{center}
\caption{Comparison between data from simulations and results from the model of section
\ref{secImpr}. The simulation data are averages over 500 realisations. Left: Mean value of the
position (with respect to the initial position), right: variance of the position. Top:
$\sigma=7\cdot 10^{-5}$ (fitted parameters: $\Theta=2.2 \cdot 10^{-4}, \tau_{A}=700$), bottom:
$\sigma=12\cdot 10^{-5}$ (fitted parameters: $\Theta=1.3 \cdot 10^{-3}, \tau_{A}=280$).}
\label{figMeanVar} \end{figure}
A comparison between the results from simulations and the
predictions of the model in section \ref{secImpr} is shown in figure \ref{figMeanVar}. We observe
that the mean value approaches a $\sigma$-dependent value constant in time, i.e. after a transient
there is no mean drift in the system. The variance, also after a transient, depends linearly on
time, which is the result expected for standard diffusion (Brownian motion). Completely
non-standard, however, is the fact that the diffusion constant, i.e. the slope of the variance as a
function of time, decreases with increasing noise strength, i.e. increasing temperature. Such a
temperature dependence has been reported for one of several diffusive processes contributing to the
diffusion of kinks in a $\phi^{4}$-model \cite{Phi4}. As we can see from figure \ref{figMeanVar},
the theoretical model developed in sections \ref{secFirst} and \ref{secImpr} reproduces this
behaviour. The agreement for the variance is quite satisfactory, for the mean value it is harder to
judge, because of the strong fluctuations still present despite the averaging over 500 realisations.
\section{A first model}
\label{secFirst}
The observation of the numerically generated trajectories reveals that the
effect the noise has on the excitation are apparently random transitions
of the configuration between basins of attraction corresponding to opposite
signs of the propagation velocity. Also, only one absolute value of the velocity plays a role; there
is no evidence of the involvement of other velocity values, which in priciple are possible in the
system.
We consider a countable set of possible velocity values $v_{i}$ together with the probabilities
$p_{i}(t)$ of the system being in the configuration corresponding to the propagation velocity
$v_{i}$ at time $t$.
From the obvious relation
\be
X(t)=\int\limits_{0}^{t} v(t^{\prime}) \td t^{\prime}
\ee
for the position $X(t)$ of the breather we obtain
\be
\langle X(t) \rangle = \int\limits_{0}^{t} \sum\limits_{i} p_{i}(t^{\prime})v_{i} \td
t^{\prime}=\sum\limits_{i}v_{i}\int\limits_{0}^{t}p_{i}(t^{\prime})\td t^{\prime}
\ee
for the mean value. For the two-time correlation function $C(t_{2},t_{1})$ we find
\begin{eqnarray}
C(t_{2},t_{1}):=\langle X(t_{2})X(t_{1})\rangle-
\langle X(t_{2})\rangle \langle X(t_{1})\rangle = \nonumber \\
= \sum\limits_{i,j} v_{i}v_{j} \int\limits_{0}^{t_{2}} \int\limits_{0}^{t_{1}}
\left[p(v_{i},t_{2}^{\prime};v_{j},t_{1}^{\prime})-p_{i}(t_{2}^{\prime})p_{j}(t_{1}^{\prime})\right]
\td t_{1}^{\prime} \td t_{2}^{\prime} = \nonumber \\
= \sum\limits_{i,j} v_{i}v_{j} \int\limits_{0}^{t_{2}} \int\limits_{0}^{t_{1}}
\left[p_{i|j}(t_{2}^{\prime}|t_{1}^{\prime})-p_{i}(t_{2}^{\prime})\right]p_{j}(t_{1}^{\prime})
\td t_{1}^{\prime} \td t_{2}^{\prime}
\end{eqnarray}
Herein $p(v_{i},t_{2}^{\prime};v_{j},t_{1}^{\prime})$ is the joint probability of finding velocity
$v_{i}$ at time $t_{2}^{\prime}$ and velocity $v_{j}$ at time $t_{1}^{\prime}$;
$p_{i|j}(t_{2}^{\prime}|t_{1}^{\prime})$ is the corresponding conditional probability
$p_{i|j}(t_{2}^{\prime}|t_{1}^{\prime})=p(v_{i},t_{2}^{\prime};v_{j},t_{1}^{\prime})/
p_{j}(t_{1}^{\prime})$.
For the case of only two velocities $+v$ and $-v$ the above specialises to
\be
\label{Mean1}
\langle X(t)\rangle = v \left\{\int\limits_{0}^{t} p_{+}(t^{\prime})\td t^{\prime} -
\int\limits_{0}^{t}\left[1-p_{+}(t^{\prime})\right]\td t^{\prime} \right\}=v\left[2\int
\limits_{0}^{t}p_{+}(t^{\prime})\td t^{\prime} - t \right],
\ee
as for only two values $p_{-}=1-p_{+}$, and
\begin{eqnarray}
\label{Corr1}
C(t_{2},t_{1})=v^{2} \int\limits_{0}^{t_{2}}\int\limits_{0}^{t_{1}}
\left\{\left[p_{+|+}(t_{2}^{\prime}|t_{1}^{\prime})-p_{+}(t_{2}^{\prime})\right]
p_{+}(t_{1}^{\prime})+
\left[p_{-|-}(t_{2}^{\prime}|t_{1}^{\prime})-p_{-}(t_{2}^{\prime})\right]
p_{-}(t_{1}^{\prime}) \right. \nonumber \\ \left.
-\left[p_{+|-}(t_{2}^{\prime}|t_{1}^{\prime})-p_{+}(t_{2}^{\prime})\right]
p_{-}(t_{1}^{\prime})
-\left[p_{-|+}(t_{2}^{\prime}|t_{1}^{\prime})-p_{-}(t_{2}^{\prime})\right]
p_{+}(t_{1}^{\prime})\right\} \td t_{1}^{\prime} \td t_{2}^{\prime}
\end{eqnarray}
Up to now nothing has been said about the probabilities occuring in \gref{Mean1} and \gref{Corr1}.
We {\it assume} that there is a constant probability $\Theta$ per unit time for a change in the sign
of the velocity. This leads to
\begin{eqnarray}
\label{ProbDiff1}
\de{p_{+}}{t}=&\Theta(p_{-}-p_{+}) \\
\de{p_{-}}{t}=&\Theta(p_{+}-p_{-})
\end{eqnarray}
with the solutions
\begin{eqnarray}
\label{Prob1}
p_{+}(t)&=\frac{1}{2}+\frac{1}{2}\left(p_{+}-p_{-}\right)(0)\exp\left(-2\Theta t\right) \\
p_{-}(t)&=\frac{1}{2}-\frac{1}{2}\left(p_{+}-p_{-}\right)(0)\exp\left(-2\Theta t\right).
\end{eqnarray}
We furthermore find:
\begin{eqnarray}
\label{Prob2}
p_{+|-}(t_{2}^{\prime}|t_{1}^{\prime})&=p_{-|+}(t_{2}^{\prime}|t_{1}^{\prime})=
\frac{1}{2}-\frac{1}{2}\exp\left[-2\Theta\left(t_{2}^{\prime}-t_{1}^{\prime}\right)\right] \\
p_{+|+}(t_{2}^{\prime}|t_{1}^{\prime})&=p_{-|-}(t_{2}^{\prime}|t_{1}^{\prime})=
\frac{1}{2}+\frac{1}{2}\exp\left[-2\Theta\left(t_{2}^{\prime}-t_{1}^{\prime}\right)\right]
\end{eqnarray}
Choosing the initial condition to have positive sign of the velocity, from these expressions and
\gref{Mean1},\gref{Corr1} we obtain
\be
\label{Mean2}
\langle X(t)\rangle =
\frac{v}{2\Theta}\left[1-\exp\left(-2\Theta t \right)\right], \qquad
\lim_{t\rightarrow \infty}\langle X(t)\rangle=
\frac{v}{2\Theta}
\ee
and
\be
C(t_{2},t_{1})=\frac{v^{2}}{2\Theta^{2}}\left[1-\exp(-2\Theta t_{2})\right]\left[\cosh(2\Theta
t_{1})-1\right]
\ee
From the latter expression we find for the variance of the position
\be
\label{Var}
Var\left[X(t)\right]=2\frac{v^{2}}{\Theta^{2}}\exp(-\Theta t)\left[\sinh(\Theta t)
\right]^{3}.
\ee
Asymptotically for long times, the variance behaves as $\exp(2\Theta t)$. This does not reflect the
behaviour found in simulations. Thus, though due to the denominators of the prefactors in
\gref{Mean2} and \gref{Var} an increase in the noise strength via an increase in $\Theta$ (this
relation between $\Theta$ and $\sigma$ is plausible and also evident from figure \ref{figTraj})
reduces
the mean value and partially has a suppressing effect on the variance, the results are not
satisfactory and an improvement of the model is called for. The next section is dedicated to it.
\section{An improved model}
\label{secImpr}
Individual trajectories obtained in numerics clearly show that the localised excitations change the
sign of their propagation velocity under the influence of the noise. This change implies a change in
the configuration of the breather, more precisely a transition from one basin of attraction to
another. Such a transition, however, does not happen instantaneously. Rather, there will be a time
span during which the configuration is not close to one of the attractors; the dynamics during
this transition period may be quite involved and is not a subject of this paper. We attempt to
roughly capture the effects of the non-instantaneousness of the transition by the introduction of a
transition time $\tau_{A}$, during which the velocity of the breather is $0$, and during which no
further transitions may be initiated. This means that if a transition from the $+v$ state is
initiated at time $t=0$, then the velocity of the localised mode immediately acquires the value
$0$, which it will retain up to $t=\tau_{A}$, when the velocity jumps to $-v$. In particular, we do
not take into consideration `failed' transitions, i.e. jumps of the velocity from $+v$ to $0$  and
back to $+v$. It is unclear whether taking into account such processes would improve the model;
reality is much more complex and can involve all kinds of trajectories in velocity space. It
certainly would increase the number of parameters and assumptions in the model; therefore we confine
ourselves to the simple model where each initiated transition after time $\tau_{A}$ ends in the
attractor corresponding to the opposite sign of the velocity. We keep the assumption of the previous
section that there is a constant probabibility $\Theta$ per unit time for a transition to be
initiated. \newline
The introduction of the `delay time' $\tau_{A}$ limits the maximum number $m$ of jumps occuring in
time $t$ to $m=[t/\tau_{A}]+1$, where $[x]$ denotes the integer part of $x$. If we consider a
trajectory with $n$ jumps up to time $t$ (note that this implies $t\geq (n-1)\tau_{A}$), the jumps
occuring at times $T_{1},T_{2},\dots,T_{n}$, with $0\leq T_{1} \leq T_{2}-\tau_{A}\leq \dots \leq
T_{n}-\tau_{A}$, we obtain for the distance traversed by the ILM:
\begin{eqnarray}
\label{X1}
X(t)=v\left[T_{1}-(T_{2}-T_{1}-\tau_{A})+(T_{3}-T_{2}-\tau_{A})-\dots \right. \nonumber \\
\left. \dots -\pi(n)(T_{n}-T_{n-1}-\tau_{A}) +\pi(n)(t-T_{n}-\tau_{A})\right]= \nonumber \\
v\left[-2\sum\limits_{k=1}^{n}\pi(k)T_{k}+\pi(n)t+\frac{1-\pi(n)}{2}\tau_{A}\right],
\end{eqnarray}
where $\pi(n)$ is the parity of $n$, i.e. $\pi(n)=+1$ if $n$ is even and $\pi(n)=-1$ if $n$ is odd.
The expression \gref{X1} holds if $T_{n} \leq t-\tau_{A}$; in the case $T_{n} > t-\tau_{A}$ we find
\be
\label{X2}
X(t)=v\left[-2\sum\limits_{k=1}^{n-1}\pi(k)T_{k}-\pi(n)T_{n}+\frac{1+\pi(n)}{2}\tau_{A}\right],
\ee
because the excitation does not move after the $n$-th jump. The temporal probability density of $n$
jumps occuring at the times $T_{1},\dots,T_{n}$ is in the case $T_{n} \leq t-\tau_{A}$:
\begin{eqnarray}
\label{IProb1} 
p(T_{n},\dots,T_{1})=\exp(-\Theta T_{1})\Theta\exp[-\Theta(T_{2}-T_{1}-\tau_{A})]\dots
\nonumber \\  \dots
\exp[-\Theta(T_{n}-T_{n-1}-\tau_{A})]\Theta\exp[-\Theta(t-T_{n}-\tau_{A})]=\Theta^{n}
\exp[-\Theta(t-n\tau_{A})]
\end{eqnarray}
and analogously in the case $T_{n} > t-\tau_{A}$:
\be
\label{IProb2}
p(T_{n},\dots,T_{1})=\Theta^{n}\exp[-\Theta(T_{n}-(n-1)\tau_{A})].
\ee
Using the equations (\ref{X1}-\ref{IProb2}) the mean value of $X(t)$ can be calculated as
\be
\label{MeanStr}
\langle X(t) \rangle = S_{0}(t)+\sum\limits_{n=1}^{m-1}\left[S_{1,n}(t)+S_{2,n}(t)\right]+S_{3,m}(t)
\ee
with
\be
S_{0}(t)=vt\exp[-\Theta t]
\ee
corresponding to no jumps, and, using the step function $H$,
\begin{eqnarray}  
S_{1,n}(t)=H(t-n\tau_{A})v\Theta^{n}\exp[-\Theta(t-n\tau_{A})]
\int\limits_{(n-1)\tau_{A}}^{t-\tau_{A}}\td T_{n}
\int\limits_{(n-2)\tau_{A}}^{T_{n}-\tau_{A}}\td T_{n-1}\dots \nonumber \\  \dots
\int\limits_{\tau_{A}}^{T_{3}-\tau_{A}}\td T_{2}
\int\limits_{0}^{T_{2}-\tau_{A}}\td T_{1}
\left[-2\sum\limits_{k=1}^{n}\pi(k)T_{k}+\pi(n)t+\frac{1-\pi(n)}{2}\tau_{A}\right],
\end{eqnarray}
\begin{eqnarray} 
S_{2,n}(t)=H(t-n\tau_{A})v\Theta^{n}
\int\limits_{t-\tau_{A}}^{t}\td T_{n} \exp[-\Theta (T_{n}-(n-1)\tau_{A})]
\int\limits_{(n-2)\tau_{A}}^{T_{n}-\tau_{A}}\td T_{n-1} \dots \nonumber \\  \dots
\int\limits_{\tau_{A}}^{T_{3}-\tau_{A}}\td T_{2}
\int\limits_{0}^{T_{2}-\tau_{A}}\td T_{1}
\left[-2\sum\limits_{k=1}^{n-1}\pi(k)T_{k}-\pi(n)T_{n}+\frac{1+\pi(n)}{2}\tau_{A}\right],
\end{eqnarray}
and
\begin{eqnarray}
S_{3,m}=H(t-(m-1)\tau_{A})\left[1-H(t-m\tau_{A})\right]v\Theta^{m} \times \nonumber \\
 \times
\int\limits_{(m-1)\tau_{A}}^{t}\td T_{m} \exp[-\Theta (T_{m}-(m-1)\tau_{A})]
\int\limits_{(m-2)\tau_{A}}^{T_{m}-\tau_{A}}\td T_{m-1} \dots
\int\limits_{\tau_{A}}^{T_{3}-\tau_{A}}\td T_{2}
\int\limits_{0}^{T_{2}-\tau_{A}}\td T_{1} \times \nonumber \\  \times
\left[-2\sum\limits_{k=1}^{m-1}\pi(k)T_{k}-\pi(m)T_{m}+\frac{1+\pi(m)}{2}\tau_{A}\right].
\end{eqnarray}
Several expressions are useful in evaluating these quantities as well as others to be introduced
below. They are gathered in an appendix. After some algebra we arrive at
\be
S_{1,n}(t)=H(t-n\tau_{A})\frac{1+\pi(n)}{2}\frac{v}{\Theta}\exp[-\Theta(t-n\tau_{A})]
\frac{\left[\Theta(t-n\tau_{A})\right]^{n+1}}{(n+1)!}
\ee
\begin{eqnarray}
S_{2,n}=H(t-n\tau_{A})\frac{1-\pi(n)}{2}\frac{v}{\Theta}\exp[-\Theta(t-n\tau_{A})] \times
\nonumber \\ \times
\sum\limits_{l=0}^{n}\frac{1}{(n-l)!}\left\{\left[\Theta(t-n\tau_{A})\right]^{n-l}-
\left[\Theta(t-(n-1)\tau_{A})\right]^{n-l}\text{e}^{-\Theta \tau_{A}}\right\}
\end{eqnarray}
\begin{eqnarray}
S_{3,m}=H(t-(m-1)\tau_{A})\left[1-H(t-m\tau_{A})\right]\frac{1-\pi(m)}{2}\frac{v}{\Theta}
\times \nonumber\\  \times
\bigg\{1-\exp\left[-\Theta(t-(m-1)\tau_{A})\right]
\sum\limits_{l=0}^{m} \frac{\left[\Theta(t-(m-1)\tau_{A})\right]^{m-l}}{(m-l)!}\bigg\}
\end{eqnarray}
The expressions for $S_{2,n}$ and $S_{3,m}$ are complicated because of the polynomials appearing; in
all the cases $S_{1,n}$,$S_{2,n}$ and $S_{3,m}$, however, we clearly see the `retardation' caused by
the introduction of $\tau_{A}$. \newline
In a similar way, the expectation of $X(t)^{2}$ can be written as
\be
\label{Mean2Str}
\langle X(t)^{2} \rangle =
Q_{0}(t)+\sum\limits_{n=1}^{m-1}\left[Q_{1,n}(t)+Q_{2,n}(t)\right]+Q_{3,m}(t)
\ee
with
\be
Q_{0}(t)=v^{2}t^{2}\exp(-\Theta t)
\ee
\begin{eqnarray}
Q_{1,n}(t)=H(t-n\tau_{A})\frac{v^{2}}{\Theta^{2}}\exp\left[-\Theta (t-n\tau_{A})\right]
\frac{\left[\Theta (t-n\tau_{A})\right]^{n+2}}{(n+2)!} \times \nonumber \\ \times
\left[\frac{1+\pi(n)}{2}(n+2)+\frac{1-\pi(n)}{2}(n+1)\right]
\end{eqnarray}
\begin{eqnarray} 
Q_{2,n}(t)=H(t-n\tau_{A})\frac{v^{2}}{\Theta^{2}}\exp\left[-\Theta(t-n\tau_{A})\right]
\sum\limits_{k=0}^{n+1}
\left\{\left[\Theta(t-n\tau_{A})\right]^{n+1-k} \right. \nonumber
\\ \left. -\left[\Theta(t-(n-1)\tau_{A})\right]^{n+1-k} \text{e}^{-\Theta\tau_{A}}\right\} \times
\nonumber \\ \times
\left[\frac{1+\pi(n)}{2}\frac{n}{(n+1-k)!}+\frac{1-\pi(n)}{2}\frac{n+1}{(n+1-k)!}\right]
\end{eqnarray}
\begin{eqnarray}
Q_{3,m}(t)=H(t-(m-1)\tau_{A})\left[1-H(t-m\tau_{A})\right]
\frac{v^{2}}{\Theta^{2}} \times \nonumber \\  \times \left\{
\Big(1-\exp\left[-\Theta(t-(m-1)\tau_{A})\right]\Big)\left(\frac{1+\pi(m)}{2}m
+\frac{1-\pi(m)}{2}(m+1)\right) \right. \nonumber \\ \left.
-\sum\limits_{k=0}^{m}
\left[\Theta(t-(m-1)\tau_{A})\right]^{m+1-k}
\exp\left[-\Theta(t-(m-1)\tau_{A})\right]\right. \times \nonumber  \\ \left.\times
\left[\frac{1+\pi(m)}{2}\frac{m}{(m+1-k)!}+\frac{1-\pi(m)}{2}\frac{m+1}{(m+1-k)!}\right]\right\}.
\end{eqnarray}
From these results we can of course obtain $Var[X(t)]=\langle X(t)^{2} \rangle -\langle X(t)
\rangle^{2}$, but as the expressions shown above are already rather involved, we do not carry out
this step explicitly. However, from the above results the mean and the variance can easily be
plotted and compared with the results from simulations, as done in figure \ref{figMeanVar}.
The results from fits for values $5\cdot 10^{-5} \leq \sigma \leq 15 \cdot 10^{-5}$ clearly show
that $\tau_{A}$ decreases with increasing noise strength (roughly as $\tau_{A}  \propto
1/(k_{B}T)$), whereas $\Theta$ increases strongly, approximately as $\Theta \propto
\exp[-\mu/(k_{B}T)]$ with some constant $\mu$. Due to the strong fluctuations, clearer results are
not possible; therefore we also do not show graphs.
\newline
The behaviour for short times $t < \tau_{A}, \; \Theta t \ll 1$,
can be obtained from $S_{3,1}$ and $Q_{3,1}$. We find
\be
\langle X(t) \rangle = \frac{v}{\Theta}\left[1-\exp\left(-\Theta t\right)\right] \approx
\frac{v}{\Theta} \left[\Theta t - \frac{1}{2} \left(\Theta t\right)^{2} \dots \right]
\ee
and
\be
Var[X(t)]=\frac{v^{2}}{\Theta^{2}}\left[1-2\Theta t \exp\left(-\Theta t\right)-\exp\left(-2\Theta
t\right) \right] \approx \frac{1}{3} \frac{v^{2}}{\Theta^{2}} \left(\Theta t\right)^{3}.
\ee
A completely different approach, which we will therefore present in a separate publication, easily
yields the long-time ($t \gg 1/\Theta$) behaviour. The results are $\langle X \rangle =
v/(2\Theta)$, which agrees with the result from section \ref{secFirst} independently of
$\tau_{A}$ and $Var[X(t)]=Dt+c$ with
\be
D=\frac{v^{2}}{\Theta^{2}\left(\tau_{A}+\frac{1}{\Theta}\right)}, \quad c \approx
-\frac{v^{2}}{2\Theta^{2}}.
\ee
Here, even in the case $\tau_{A} \rightarrow 0$, an agreement with the simpler model of the previous
section, where the variance shows exponential behaviour, is not achieved.
\section{Discussion and Outlook}
\label{secDiss}
As we have seen, the diffusive behaviour of an intrinsic localised mode in the damped-driven
discrete sine-Gordon chain, when Gaussian white noise is coupled to the system to model the effects
of temperature, is due to random transitions between attracting configurations of the system that
correspond to opposite signs of the propagation velocity of the mode.
In this paper we have developed a model that satisfactorily describes this diffusion process. At a
preliminary stage, the model only uses the transition probability per unit time, $\Theta$, as
parameter. This approach is not sufficient; it is improved by the introduction of a second
parameter, the delay time $\tau_{A}$. Both parameters are determined from fits to the results from
simulations.
These fits show that with increasing temperature $\tau_{A}$ decreases and $\Theta$ increases.
The long time behaviour for both models is the same for the expectation of the position
of the intrinsic localised mode, but not for the variance of the position, not even in the case of
vanishing delay time. This is due to a fundamental difference between the two approaches we have
presented. In the one-parameter model we have only taken into account the probabilities of finding
the configurations in one of the attracting states or the other. The two-parameter model, on the
other hand, follows single trajectories through time and considers the probabilities for
repeated jumps between the basins of attraction to occur in one and the same trajectory. The
nonvanishing delay time, during which propagation of the excitation and further jumps are forbidden,
also introduces a non-Markovian element in the diffusion process.
The model does not make use of any specific characteristics of the
system; it only requires the existence of the attracting states and the possibility to excite
transitions between these states by noise. The details of such a transition will, we suspect, be
sensitive to the particular system, but in the model that we have presented in section \ref{secImpr}
the details are `hidden' in the delay time $\tau_{A}$. Thus the origin of deviations between the
prediction of the model and the result of simulations, which show in the transient period between
the very early stages of the stochastic evolution and the long time diffusive regime, lies in the
dynamics of the transition from one attractor to the other. \newline
A first step in the development of a more detailed model would be a calculation of the parameters
$\Theta$ and $\tau_{A}$ from the system parameters $C, \alpha, F, \omega_{0}$ and the noise strength
$\sigma$. A further step would be to go beyond the simple picture of `delayed jumps' that is at
the heart of our model and to include the dynamics in full. The latter step then should also be able
to describe the behaviour in the transient time regime.
\section*{Acknowledgements}
MM and LV acknowledge support from the European Commission within the Research Training
Network (RTN) LOCNET, contract HPRN-CT-1999-00163. LV also was partially supported by the
Ministry of Science and Technology of Spain through grant BFM2002-02345.
\appendix
\section*{Appendix}
\setcounter{section}{1}
We list here several expressions that arise in the evaluation of the mean value of $X$ and of
$X^{2}$. $M$ in the equations below is not to be confused with the maximum number of jumps $m$.
\be
\int\limits_{(M-2)\tau_{A}}^{T_{M}-\tau_{A}}\td T_{M-1}
\int\limits_{(M-3)\tau_{A}}^{T_{M-1}-\tau_{A}}\td T_{M-2} \dots
\int\limits_{\tau_{A}}^{T_{3}-\tau_{A}}\td T_{2}
\int\limits_{0}^{T_{2}-\tau_{A}}\td T_{1}=
\frac{\left[T_{M}-(M-1)\tau_{A}\right]^{M-1}}{(M-1)!}
\ee
For $1\leq k \leq M-1$
\begin{eqnarray}
\int\limits_{(M-2)\tau_{A}}^{T_{M}-\tau_{A}}\td T_{M-1}
\int\limits_{(M-3)\tau_{A}}^{T_{M-1}-\tau_{A}}\td T_{M-2} \dots
\int\limits_{\tau_{A}}^{T_{3}-\tau_{A}}\td T_{2}
\int\limits_{0}^{T_{2}-\tau_{A}}\td T_{1}T_{k}= \nonumber \\
k\frac{\left[T_{M}-(M-1)\tau_{A}\right]^{M}}{M!}+(k-1)\tau_{A}
\frac{\left[T_{M}-(M-1)\tau_{A}\right]^{M-1}}{(M-1)!}
\end{eqnarray}
\begin{eqnarray}
\int\limits_{(M-2)\tau_{A}}^{T_{M}-\tau_{A}}\td T_{M-1}
\int\limits_{(M-3)\tau_{A}}^{T_{M-1}-\tau_{A}}\td T_{M-2} \dots
\int\limits_{\tau_{A}}^{T_{3}-\tau_{A}}\td T_{2}
\int\limits_{0}^{T_{2}-\tau_{A}}\td T_{1}T_{k}^{2}= \nonumber \\
k(k+1)\frac{\left[T_{M}-(M-1)\tau_{A}\right]^{M+1}}{(M+1)!}+
2\tau_{A}k(k-1)\frac{\left[T_{M}-(M-1)\tau_{A}\right]^{M}}{M!}+  \nonumber \\
(k-1)^{2}\tau_{A}^{2}\frac{\left[T_{M}-(M-1)\tau_{A}\right]^{M-1}}{(M-1)!}
\end{eqnarray}
For $1\leq k < l \leq M-1$
\begin{eqnarray}
\int\limits_{(M-2)\tau_{A}}^{T_{M}-\tau_{A}}\td T_{M-1}
\int\limits_{(M-3)\tau_{A}}^{T_{M-1}-\tau_{A}}\td T_{M-2} \dots
\int\limits_{\tau_{A}}^{T_{3}-\tau_{A}}\td T_{2}
\int\limits_{0}^{T_{2}-\tau_{A}}\td T_{1}T_{k}T_{l}= \nonumber \\
(kl+k)\frac{\left[T_{M}-(M-1)\tau_{A}\right]^{M+1}}{(M+1)!}+
(2kl-l-k)\tau_{A}\frac{\left[T_{M}-(M-1)\tau_{A}\right]^{M}}{M!}+ \nonumber \\
(kl-l-k+1)\tau_{A}^{2}\frac{\left[T_{M}-(M-1)\tau_{A}\right]^{M-1}}{(M-1)!}
\end{eqnarray}
\be
\sum\limits_{k=1}^{n}\pi(k)=-\frac{1-\pi(n)}{2}
\ee
\be
\sum\limits_{k=1}^{n}\pi(k)k=\frac{1+\pi(n)}{2}\frac{n}{2}+\frac{1-\pi(n)}{2}\left(-\frac{n+1}{2}
\right)
\ee
\begin{eqnarray}
\sum\limits_{l=k+1}^{n}\pi(l)l=\frac{1+\pi(n)}{2}\frac{n}{2}+\frac{1-\pi(n)}{2}\left(-\frac{n+1}{2}
\right)-  \nonumber \\
\frac{1+\pi(k)}{2}\frac{k}{2}-\frac{1-\pi(k)}{2}\left(-\frac{k+1}{2}
\right)
\end{eqnarray}
\be
\sum\limits_{k=1}^{n-1}\pi(k)\sum\limits_{l=k+1}^{n}\pi(l)l=
\frac{1+\pi(n)}{2}\left(-\frac{n^{2}}{4}-\frac{n}{2}\right)+
\frac{1-\pi(n)}{2}\left(\frac{1}{4}-\frac{n^{2}}{4}\right)
\ee
\be
\sum\limits_{k=1}^{n-1}\pi(k)\sum\limits_{l=k+1}^{n}\pi(l)=
-\frac{1}{2}\frac{1+\pi(n)}{2}-\frac{1}{2}(n-1)
\ee
\be
\sum\limits_{k=1}^{n-1}\pi(k)k\sum\limits_{l=k+1}^{n}\pi(l)=
\frac{1-\pi(n)}{2}\left(-\frac{n^{2}}{4}+\frac{1}{4}\right)+\frac{1+\pi(n)}{2}(-\frac{n^{2}}{4})
\ee
\begin{eqnarray}
\sum\limits_{k=1}^{n-1}\pi(k)\sum\limits_{l=k+1}^{n}\pi(l)kl=
\frac{1+\pi(n)}{2}\left(-\frac{n^{3}}{6}-\frac{n^{2}}{8}-\frac{n}{12}\right)+ \nonumber \\
\frac{1-\pi(n)}{2}\left(-\frac{n^{3}}{6}-\frac{n^{2}}{8}+\frac{n}{6}+\frac{1}{8}\right)
\end{eqnarray}


\begin{thebibliography}{00}
\bibitem{DisNon} Peyrard M 1998 {\it Physica} D {\bf 123} 403
\bibitem{MaAu94} MacKay R S and Aubry S 1994 {\it Nonlinearity} {\bf 7} 1623
\bibitem{Au97} Aubry S 1997 {\it Physica} D {\bf 103} 201
\bibitem{Sep97} Sepulchre J-A and MacKay R S 1997 {\it Nonlinearity} {\bf 10} 679
\bibitem{Mart03} Mart\'\i nez P J, Meister M, Flor\'\i a L M and Falo F 2003 {\it Chaos} to appear
\bibitem{Siev88} Sievers A J and Takeno S 1988 {\it Phys. Rev. Lett.} {\bf 61} 970
\bibitem{Tak90} Takeno S and Hori K 1990 {\it J. Phys. Soc. Japan} {\bf 59} 3037
\bibitem{Cuev02-1} Cuevas J, Palmero F, Archilla J F R and Romero F R 2002 {\it J. Phys. A: Math.
Gen.} {\bf 35} 10519
\bibitem{Cuev02-2} Cuevas J, Palmero F, Archilla J F R and Romero F R 2002 {\it
Phys. Lett.} A {\bf 299} 221
\bibitem{Cuev02-3} Cuevas J, Archilla J F R, Gaididei Yu B and Romero F R 2002 {\it Physica} D {\bf
163} 106
\bibitem{Pas85} Pascual P J and V\'azquez L 1985 {\it Phys. Rev.} B {\bf 32} 8305
\bibitem{Rod90} Rodr\'\i guez-Plaza M J and V\'azquez L 1990 {\it Phys. Rev.} B {\bf 41} 11437
\bibitem{Phi4} Ivanov B A and Kolezhuk A K 1990 {\it Phys. Lett.} A {\bf 146} 190
\bibitem{Qui99} Quintero N R, S\'anchez A and Mertens F G 1999 {\it Phys. Rev.} E {\bf 60} 222
\bibitem{Qui00} Quintero N R, S\'anchez A and Mertens F G 2000 {\it Eur. Phys. J.} B {\bf 16} 361
\bibitem{Mei01} Meister M, Mertens F G and S\'anchez A 2001 {\it Eur. Phys. J.} B {\bf 20} 405
\bibitem{JJA} Ustinov A V 1998 {\it Physica} D {\bf 123} 315
\bibitem{Mar01}
Mar\'\i n J L, Falo F, Mart\'\i nez P J and Flor\'\i a L M  2001 {\it Phys. Rev.} E {\bf 63} 066603
\bibitem{Gard} Gardiner C W  1985 {\it Handbook of Stochastic Methods} (Berlin: Springer)
\end{thebibliography}
\end{document}